# Target Defense Using a Turret and Mobile Defender Team

Alexander Von Moll[1], Dipankar Maity[2], Meir Pachter[3], Daigo Shishika[4], and Michael Dorothy[5]

*Abstract*—A scenario is considered wherein a stationary, turn constrained agent (Turret) and a mobile agent (Defender) cooperate to protect the former from an adversarial mobile agent (Attacker). The Attacker wishes to reach the Turret prior to getting captured by either the Defender or Turret, if possible. Meanwhile, the Defender and Turret seek to capture the Attacker as far from the Turret as possible. This scenario is formulated as a differential game and solved using a geometric approach. Necessary and sufficient conditions for the Turret-Defender team winning and the Attacker winning are given. In the case of the Turret-Defender team winning equilibrium strategies for the $\min \max$ terminal distance of the Attacker to the Turret are given. Three cases arise corresponding to solo capture by the Defender, solo capture by the Turret, and capture simultaneously by both Turret and Defender.

## I. INTRODUCTION

Point defense is a critical task for securing an area from mobile threats [1]. It involves identifying the threat(s) as well as allocating defensive resources to neutralize them. The task is especially pertinent based on recent attacks involving more than 300 airborne threats aimed at ground-based targets [2]. As indicated by [3], there is a need for coordination among different types of defensive assets, both ground-based and airborne, to achieve success.

This paper considers a scenario in which a turret and mobile defender cooperate to defend a circular region against a single incoming threat (see Fig. 1).

This paper is based on work performed at the Air Force Research Laboratory (AFRL) *Control Science Center* and supported by AFOSR LRIR #24RQCOR002 (funded by Dr. Frederick Leve) and DEVCOM ARL grant ARL DCIST CRA W911NF-17-2-0181. DISTRIBUTION STATEMENT A. Approved for public release. Distribution is unlimited. AFRL-2025-0001003; Cleared 28 Aug 2025.

[1]Von Moll is with Control Science Center, Air Force Research Laboratory, 2210 8th St, WPAFB, OH, USA, alexander.von_moll@afrl.af.mil.

[2]Maity is with Department of Electrical and Computer Engineering, University of North Carolina at Charlotte, 9201 University City Blvd, Charlotte, NC, USA, dmaity@charlotte.edu.

[3]Pachter is with Department of Electrical Engineering, Air Force Institute of Technology, 2950 Hobson Way, WPAFB, OH, USA, meir.pachter@afit.edu.

[4]Shishika is with Department of Mechanical Engineering, George Mason University, 4400 University Dr, Fairfax, VA, USA, dshishik@gmu.edu.

[5]Dorothy is with DEVCOM, Army Research Laboratory, 2800 Powder Mill Rd, Adelphi, MD, USA, michael.r.dorothy.civ@army.mil.

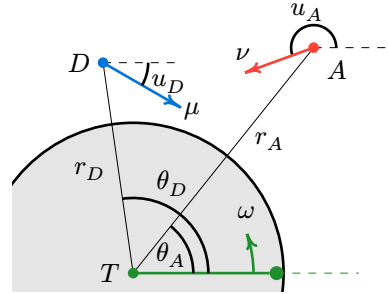

Fig. 1. Turret defense scenario.

Agent names are capitalized (after the style of Isaacs [4]). The Turret is modeled as a stationary agent which can turn with a bounded turn rate. The Defender and Attacker are modeled as constant speed agents with control over their instantaneous heading (i.e., have simple motion, are holonomic). The target is a circle centered on the Turret's position. The Turret-Defender team is successful if it can capture the Attacker prior to the latter reaching the target region. In order to capture the Attacker, either the Defender must collide with it or the Turret must align to its position. Thus the main objective of this work is to analyze the cooperation among this team comprised of heterogeneous agents. There are many considerations that must be made for large-scale cooperation among heterogeneous robotic teams such as task decomposition, task allocation, sensing, and motion control [5]. This study focuses on the motion control aspect (essentially assuming all preceding items in the list have been addressed *a priori*). The interesting and challenging problem of task allocation (e.g., in the case of a many-versus-many engagement) is left for follow-on studies, though several approaches have proven to be effective (c.f., e.g., [6]). This work poses the scenario as a differential game [4] in order to obtain equilibrium strategies for both sides simultaneously. Differential game theory has previously proven useful in the analysis of turret defense scenarios [7].

For guidance applications, in particular pursuit-evasion scenarios, there are several examples of cooperation among a team of homogeneous vehicles. In the seminal work [8], two slow pursuers cooperate to minimize their approach distance to an evader who must pass between them. By designing the control inputs for both pursuers simultaneously they can achieve better performance than if they were to act in isolation. This is an example of explicit cooperation. Similarly, both [9] and [10] analyzed multiple-pursuer, single-evader scenarios although with an objective of $\min \max$ capture time. The presence of additional pursuers further limited the survival time the evader could achieve.

There are also examples of cooperation among teams of like vehicles but with heterogeneous roles. A well-known example is the active target defense scenario wherein a target and defender cooperate to protect the target from an attacker [11]. Another is a perimeter defense example with patrollers (who have a preset route) and defenders (who react to incoming attackers) [12]. In [13] two mobile attackers cooperate to reach a turret by having one attacker serve as a sacrifice in order for the other to survive.

This work makes several contributions: 1) the Attacker's dominance region w.r.t. the Turret is characterized for the first time 2) equilibrium strategies and Value function are obtained for the game of min max terminal Attacker distance 3) closed-form analytic conditions are given for the determination of how the game ends. Importantly, this work demonstrates how differential game theory can address cooperation among a team with completely different capabilities.

The remainder of this paper is organized as follows. Section II provides the mathematical formulation of the scenario. Section III utilizes the necessary conditions for equilibrium to ascertain properties of the equilibrium control strategies. Section IV focuses on the geometry of the Attacker's dominance regions w.r.t. the Turret and Defender. Section V contains the main results and the paper is concluded in Section VI.

## II. SCENARIO FORMULATION

Let the boundary of the target region be specified by a circle centered on the Turret's position whose radius, without loss of generality, is set to 1. Let $\nu, \mu$, and $\omega$ be the linear speed of the Attacker, linear speed of the Defender, and maximum turn rate of the Turret, respectively. It is assumed that the Attacker is slower than the Defender and Turret, i.e., $\nu < \mu$ and $\nu < \omega$. $T$ is assumed to be at the origin w.l.o.g with initial look angle, $\theta_T$, aligned with the positive $x$-axis. The kinematics are

$$\mathbf{f}(\mathbf{x}, u_D, u_A, u_T) = \dot{\mathbf{x}} = \begin{bmatrix} \dot{x}_D \\ \dot{y}_D \\ \dot{x}_A \\ \dot{y}_A \\ \dot{\theta}_T \end{bmatrix} = \begin{bmatrix} \mu \cos u_D \\ \mu \sin u_D \\ \nu \cos u_A \\ \nu \sin u_A \\ \omega u_T \end{bmatrix}, \quad (1)$$

$$u_D, u_A \in [-\pi, \pi], \quad u_T \in [-1, 1].$$

where the subscripts $D, A, T$ correspond to the Defender, Attacker, and Turret, respectively. Also, define the positions of the Defender and Attacker as $D = [x_D, y_D]^\top$ and $A = [x_A, y_A]^\top$, respectively. For notational convenience, define the polar coordinates for a point $P$ w.r.t. $T$:

$$r_P = \|P\|, \quad \theta_P = \operatorname{atan2}(y_P, x_P) - \theta_T, \quad (2)$$

where $\|\cdot\|$ represents the 2-norm.

Let capture by the Defender and Turret be defined, respectively, as

$$\begin{aligned} \mathcal{C}_D &= \{\mathbf{x} \mid r_D = r_A > 1 \text{ and } \theta_D = \theta_A\}, \\ \mathcal{C}_T &= \{\mathbf{x} \mid \theta_A = 0 \text{ and } r_A > 1\}. \end{aligned} \quad (3)$$

Thus the set $\mathcal{C}_{D,T} = \mathcal{C}_D \cap \mathcal{C}_T$ represents simultaneous capture by both $D$ and $T$.

Consider the case wherein the $D$-$T$ team is able to capture $A$ before the latter can reach its goal of $r_A = 1$. Let the final time be defined as

$$t_f = \inf\{t \mid \mathbf{x} \in \mathcal{C}_D \cup \mathcal{C}_T\}, \quad (4)$$

and the final state as $\mathbf{x}_f = \mathbf{x}(t_f)$. The goal of $A$ is to get as close as possible to its goal whereas the $D$-$T$ team seeks to capture $A$ as far from $T$ as possible. Thus the objective cost functional is defined as

$$J(\mathbf{x}, u_D(\cdot), u_A(\cdot), u_T(\cdot)) = r_{A_f} - 1, \quad (5)$$

which $A$ aims to minimize and the $D$-$T$ team wishes to maximize. This scenario falls under the category of a two-player, zero-sum differential game. The Value function, if it exists, satisfies *Isaacs' condition* [4]

$$\begin{aligned} V(\mathbf{x}) &= \min_{u_A(\cdot)} \max_{u_D(\cdot), u_T(\cdot)} J(\mathbf{x}, u_D(\cdot), u_A(\cdot), u, u_T(\cdot)) \\ &= \max_{u_D(\cdot), u_T(\cdot)} \min_{u_A(\cdot)} J(\mathbf{x}, u_D(\cdot), u_A(\cdot), u, u_T(\cdot)). \end{aligned} \quad (6)$$

This Value function also satisfies the Nash equilibrium property:

$$J(\mathbf{x}, u_D, u_A^*, u_T) \leq \underbrace{J(\mathbf{x}, u_D^*, u_A^*, u_T^*)}_{V(\mathbf{x})} \leq J(\mathbf{x}, u_D^*, u_A, u_T^*) \quad (7)$$

$$\forall u_D \in \mathcal{U}_D, u_A \in \mathcal{U}_A, u_T \in \mathcal{U}_T,$$

where $\mathcal{U}_D, \mathcal{U}_A$, and $\mathcal{U}_T$ are the sets of admissible controls for the agents, respectively. The starred $(*)$ strategies are referred to as the equilibrium strategies.

## III. EQUILIBRIUM STRATEGIES

In order to characterize the optimal/equilibrium strategies the necessary conditions for equilibrium are utilized. The Hamiltonian is formed as the inner product of a vector of costates, $\boldsymbol{\lambda} = \begin{bmatrix} \lambda_{x_A}, \lambda_{y_A}, \lambda_{x_D}, \lambda_{y_D}, \lambda_{\theta_T} \end{bmatrix}^\top$, with the kinematics (since this is a terminal cost game):

$$\begin{aligned} \mathcal{H} &= \lambda_{x_A} \nu \cos u_A + \lambda_{y_A} \nu \sin u_A \\ &+ \lambda_{x_D} \mu \cos u_D + \lambda_{y_D} \mu \sin u_D + \lambda_{\theta_T} \omega u_T. \end{aligned} \quad (8)$$

The necessary conditions for equilibrium dictate that the equilibrium costate dynamics satisfy $\dot{\boldsymbol{\lambda}} = -\frac{\partial \mathcal{H}}{\partial \mathbf{x}}$ [14] which implies

$$\dot{\boldsymbol{\lambda}} = 0. \quad (9)$$

Another necessary condition for equilibrium is that the equilibrium controls for $A$ and the $D$-$T$ team must minimize and maximize, respectively, the Hamiltonian:

$$\cos u_A^* = -\frac{\lambda_{x_A}}{\sqrt{\lambda_{x_A}^2 + \lambda_{y_A}^2}}, \quad \sin u_A^* = -\frac{\lambda_{y_A}}{\sqrt{\lambda_{x_A}^2 + \lambda_{y_A}^2}} \quad (10)$$

$$\cos u_D^* = \frac{\lambda_{x_D}}{\sqrt{\lambda_{x_D}^2+\lambda_{y_D}^2}}, \quad \sin u_D^* = \frac{\lambda_{y_D}}{\sqrt{\lambda_{x_D}^2+\lambda_{y_D}^2}} \quad (11)$$

$$u_T^* = \text{sign}(\lambda_{\theta_T}) \quad (12)$$

**Lemma 1.** *For the differential game specified by (1)–(6) the Attacker and Defender's equilibrium strategies are to take a straight-line trajectory while the Turret's equilibrium strategy is to turn in one direction with its max turn rate.*

*Proof.* The result follows from the fact that the equilibrium control strategies, (10)–(12), are functions only of the costates which are constant. ∎

## IV. CAPTURE GEOMETRIES

In this section, the three termination cases for the Attacker-losing scenario are analyzed in detail. For the cases of solo capture (by $D$ or by $T$ alone), the Attacker's dominance region w.r.t. the capturing agent is derived. Additionally, differential games involving the $A$ and the capturing agent are solved and connected to the geometry of the Attacker's respective dominance region.

### A. Capture by the Defender

In the first termination case, $A$ is captured by $D$ alone; the engagement ends with the agents being coincident per (3). This situation may arise when $T$ is initially pointed far away from $A$ and $D$ or if $T$ turns relatively slowly, for example. For all of the analysis in this section, $T$ is ignored and its influence on the kinematics is zeroed out. When the Turret is not involved, the scenario is a particular case of *Guarding a Target* from [4].

First define $A$'s dominance region w.r.t. $D$ as the set of all points in which $A$ can reach before $D$ can:

$$\mathcal{R}_{A/D} = \left\{ \mathbf{y} \in \mathbb{R}^2 \ \bigg| \ \frac{\|\mathbf{y} - A\|}{\nu} < \frac{\|\mathbf{y} - D\|}{\mu} \right\}. \quad (13)$$

It is well known (c.f., e.g., [4]) that, under the simple motion model in (1), the boundary of this region is given by the Apollonius circle, i.e., the circle whose center and radius are given by

$$\mathbf{c} = (x_\mathbf{c}, y_\mathbf{c}) = (1+\alpha)A - \alpha D, \quad \rho = \frac{\mu\alpha}{\nu}\|D - A\|, \quad (14)$$

where $\alpha = \frac{\nu^2}{\mu^2 - \nu^2}$. Thus $\mathcal{R}_{A/D}$ can be redefined as

$$\mathcal{R}_{A/D} = \{\mathbf{y} \in \mathbb{R}^2 \mid \|\mathbf{y} - \mathbf{c}\| < \rho\}. \quad (15)$$

It is assumed that the origin is not contained within $\mathcal{R}_{A/D}$.

**Lemma 2.** *For the differential game specified by (1) – (6), when $T$ has no effect on the outcome of the game under equilibrium play, the equilibrium capture point is the closest point on the Apollonius circle, $\partial \mathcal{R}_{A/D}$, to $T$. That is the Value of the game is*

$$V(\mathbf{x}) = \|\mathbf{c}\| - \rho - 1. \quad (16)$$

*The equilibrium strategies of $A$ and $D$ are given by*

$$\cos u_A^* = \frac{x_{\mathbf{p}_D}^* - x_A}{\|\mathbf{p}_D^* - A\|}, \quad \sin u_A^* = \frac{y_{\mathbf{p}_D}^* - y_A}{\|\mathbf{p}_D^* - A\|}$$
$$\cos u_D^* = \frac{x_{\mathbf{p}_D}^* - x_D}{\|\mathbf{p}_D^* - D\|}, \quad \sin u_D^* = \frac{y_{\mathbf{p}_D}^* - y_D}{\|\mathbf{p}_D^* - D\|}, \quad (17)$$

*where the equilibrium intercept point is defined as*

$$\mathbf{p}_D^* = (x_{\mathbf{p}_D}^*, y_{\mathbf{p}_D}^*) = \frac{\mathbf{c}}{\|\mathbf{c}\|}(\|\mathbf{c}\| - \rho). \quad (18)$$

*Proof.* Since it is assumed that $T$ has no effect on the outcome of the game its relevant costate $\lambda_\theta$ is set to 0. Substituting the equilibrium controls, (10) and (11), into the Hamiltonian, (8), yields

$$\mathcal{H} = -\nu\sqrt{\lambda_{x_A}^2 + \lambda_{y_A}^2} + \mu\sqrt{\lambda_{x_D}^2 + \lambda_{y_D}^2} \quad (19)$$

According to Isaacs' Verification Theorem [4], a candidate Value function may be verified as the solution to a differential game by ensuring it is continuously differentiable and satisfies the Hamilton-Jacobi-Isaacs (HJI) equation:

$$\min_{u_A} \max_{u_D} \left\{ l(\mathbf{x}, u_A, u_D) + \frac{\partial V}{\partial t} + \nabla_\mathbf{x} V \cdot \dot{\mathbf{x}}^* \right\} = 0, \quad (20)$$

where $l$ is a running cost, which is 0 in this case. The candidate Value function stated above does not depend on time and thus $\frac{\partial V}{\partial t} = 0$. Note that, by construction, $\boldsymbol{\lambda} = \nabla_\mathbf{x} V$ and therefore the Hamiltonian is identical to the term $\nabla_\mathbf{x} V \cdot \dot{\mathbf{x}}^*$. Therefore, it is sufficient to show that $\mathcal{H} = 0$. Taking partial derivatives of (16) w.r.t. the states gives

$$\lambda_{x_A} = \frac{\partial V}{\partial x_A} = \frac{((1+\alpha)x_A - \alpha x_D)(1+\alpha)}{\|\mathbf{c}\|} - \frac{\mu\alpha}{\nu}\frac{x_A - x_D}{\|D-A\|}$$
$$\lambda_{y_A} = \frac{\partial V}{\partial y_A} = \frac{((1+\alpha)y_A - \alpha y_D)(1+\alpha)}{\|\mathbf{c}\|} - \frac{\mu\alpha}{\nu}\frac{y_A - y_D}{\|D-A\|}$$
$$\lambda_{x_D} = \frac{\partial V}{\partial x_D} = -\frac{((1+\alpha)x_A - \alpha x_D)\alpha}{\|\mathbf{c}\|} + \frac{\mu\alpha}{\nu}\frac{x_A - x_D}{\|D-A\|} \quad (21)$$
$$\lambda_{y_D} = \frac{\partial V}{\partial y_D} = -\frac{((1+\alpha)y_A - \alpha y_D)\alpha}{\|\mathbf{c}\|} + \frac{\mu\alpha}{\nu}\frac{y_A - y_D}{\|D-A\|}.$$

It is straight forward to show that substituting the above expressions into (19) yields $\mathcal{H} = 0$ as required. The equilibrium control strategies, (17), follow from Lemma 1 and the fact that the closest point on the Apollonius circle to the origin lies on the line passing through its center. ∎

This result is a special case of the more general result from [15] which accounts for arbitrary convex target sets, multiple pursuers, and higher dimensions.

### B. Capture by the Turret

In the second termination case, $A$ is captured by $T$ alone; the engagement ends with $T$ aligning its line-of-sight with $A$'s position, i.e. driving $\theta_A \to 0$ per (3). This situation may arise if, for example, $T$ is initially aimed near $A$ or $D$ is too far away to affect the outcome. As before, the states pertaining to $D$, who is redundant in this case, are ignored. For the analysis in this section, without loss of generality, let $\theta_A \in (0, 2\pi)$ and assume that $T$ turns CCW (i.e., $u_T = 1$).

Let $\mathcal{P} = (0, \infty) \times (-\pi, \pi]$ be the set of polar coordinates with angles measured w.r.t. $T$'s look angle. Define $A$'s dominance region w.r.t. $T$ as the set of all points in which $A$ can reach before $T$ can align with it.

**Lemma 3.** *The dominance region of $A$ w.r.t. $T$ is given by*
$$\mathcal{R}_{A/T} = \big\{(r, \theta) \in \mathcal{P} \;\big|\; \\ r^2 + r_A^2 - 2rr_A \cos(\theta - \theta_A) \leq \tfrac{\nu^2 \theta^2}{\omega^2}\big\}. \tag{22}$$

*Proof.* The fastest way for $A$ to reach the point $\mathbf{p} = (r, \theta)$ is to take a straight-line trajectory. Consider the triangle $\triangle TA\mathbf{p}$ whose sides are given by $r_A$, $r$, and $A$'s path to $\mathbf{p}$. The Law of Cosines based on the angle at $T$ yields
$$\|\mathbf{p} - A\|^2 = r^2 + r_A^2 - 2rr_A \cos(\theta - \theta_A). \tag{23}$$
$A$'s squared time-to-go is thus $\|\mathbf{p} - A\|^2/\nu^2$. Meanwhile, $T$'s squared time-to-go is based on turning directly to $\theta$, i.e., $\theta^2/\omega^2$. Hence (22) follows. ∎

Consider the boundary of $A$'s dominance region w.r.t. $T$, i.e., $\partial \mathcal{R}_{A/T}$ which is obtained when the equality of (22) holds. Solving for $r$ gives
$$r = r_A \cos(\theta - \theta_A) \pm \sqrt{\tfrac{\nu^2 \theta^2}{\omega^2} - r_A^2 \sin^2(\theta - \theta_A)}. \tag{24}$$
Thus, in order for $r$ to have two non-negative real roots the following must be satisfied
$$\theta \in \left[-\tfrac{\pi}{2} + \theta_A, \tfrac{\pi}{2} + \theta_A\right], \tag{25}$$
$$r_A \geq \tfrac{\nu|\theta|}{\omega} \geq r_A |\sin(\theta - \theta_A)|. \tag{26}$$
The left-hand inequality of (26) implies that $T$ must be able to turn an angle $\theta$ at or before the time in which $A$ can traverse a distance $r_A$. For the remainder, it is assumed that the LHS is satisfied, i.e., $A$ starts sufficiently far from $T$. Thus the boundaries of the $\theta$ domain of $\mathcal{R}_{A/T}$ are defined by the points where the RHS inequality of (26) is equal. The following results are used to obtain these boundaries. Define a function
$$g(r) = \sqrt{\tfrac{\omega^2 r^2}{\nu^2} - 1} + \sin^{-1}\left(\tfrac{\nu}{r\omega}\right). \tag{27}$$

**Lemma 4.** *If the state of $A$ is such that $\theta_A < \theta_B(r_A)$ where*
$$\theta_B(r) = \sqrt{\tfrac{\omega^2 r^2}{\nu^2} - 1} - \cos^{-1}\left(\tfrac{\nu}{r\omega}\right) \tag{28}$$
*then $A$ cannot guarantee being able to reach $r \leq \tfrac{\nu}{\omega}$ without getting captured by $T$.*

*Proof.* From [16] the barrier which separates $A$ being able to reach a distance $\tfrac{\nu}{\omega}$ to $T$ or not is given by $\theta_B(r) = g(r) - g\big(\tfrac{\nu}{\omega}\big)$. Substituting in (27) and evaluating at $r = r_A$ yields the result. ∎

**Lemma 5.** *Consider $\theta_A > 0$ without loss of generality. If the state of $A$ is such that $\theta_A > \theta_B(r_A)$ then: 1) there exists exactly one angle, $\underline{\theta} \in (\max\{0, \theta_A - \tfrac{\pi}{2}\}, \theta_A]$, and exactly one angle, $\overline{\theta} \in (\theta_A, \theta_u]$, such that*
$$\tfrac{\nu \theta}{\omega} = r_A |\sin(\theta - \theta_A)|, \tag{29}$$
*where*
$$\theta_u = \cos^{-1}\left(\tfrac{\nu}{\omega r_A}\right) + \theta_A \tag{30}$$
*and 2) the range $[\underline{\theta}, \overline{\theta}]$ is the domain of $A$'s dominance region w.r.t. $T$ as given by (24).*

*Proof.* For convenience, define $m(\theta) = \tfrac{\nu \theta}{\omega}$ and $n(\theta) = r_A |\sin(\theta - \theta_A)|$; the derivatives of these functions w.r.t. $\theta$ are denoted as $m'$ and $n'$, respectively:
$$\begin{aligned} m'(\theta) &= \tfrac{\nu}{\omega} \\ n'(\theta) &= r_A \operatorname{sign}(\sin(\theta - \theta_A)) \cos(\theta - \theta_A) \end{aligned} \tag{31}$$
First, define $\theta_{\min} = \max\{0, \theta_A - \tfrac{\pi}{2}\}$ and consider the range $\theta \in (\theta_{\min}, \theta_A]$. The function $n(\theta)$ is monotonically decreasing while the function $m(\theta)$ is monotonically increasing. Also, $n(\theta_{\min}) > m(\theta_{\min})$ and $n(\theta_A) < m(\theta_A)$, which implies there is a unique angle in this range, $\underline{\theta}$, where $m(\underline{\theta}) = n(\underline{\theta})$. Next, consider the range $\theta \in (\theta_A, \theta_u]$. In this range, $\operatorname{sign}(\sin(\theta - \theta_A)) = 1$ since $\theta_u < \theta_A + \tfrac{\pi}{2}$, so the derivative of $n$ simplifies to
$$n'(\theta) = r_A \cos(\theta - \theta_A). \tag{32}$$
Both $m$ and $n$ are monotonically increasing in this range (i.e., $m', n' > 0$). However, $n'' = -r_A \sin(\theta - \theta_A) < 0$ so $n'$ is monotonically decreasing in this range. At $\theta = \theta_A$, we have $n'(\theta_A) > m'(\theta_A)$ (i.e., $r_A > \tfrac{\nu}{\omega}$ since, otherwise, $A$ has angular rate advantage over $T$). By construction, at $\theta_u$, we have $n'(\theta_u) = m'(\theta_u) = \tfrac{\nu}{\omega}$. Therefore, due to the monotonicity of $n'$, we have $n' > m'$ in the whole range. Thus, if an intersection exists in this range, it must be unique. Finally, if $n(\theta_u) > m(\theta_u)$, then an intersection must exist due to the Intermediate Value Theorem. This condition is $r_A \sin(\theta_u - \theta_A) > \tfrac{\nu \theta_u}{\omega}$. After substituting in (30) and rearranging, this condition becomes $\theta_A \leq \theta_B(r_A)$ which was one of the premises. Therefore $A$'s dominance region w.r.t. $T$ is well-defined over the domain $[\underline{\theta}, \overline{\theta}]$. ∎

**Lemma 6.** *For the differential game specified by (1) – (6), when $D$ has no effect on the outcome of the game under equilibrium play, the Value of the game is*
$$V(\mathbf{x}) = g^{-1}(g(r_A) - \theta_A) - 1 \tag{33}$$
*and the equilibrium strategy for $A$ is*
$$\hat{u}_A^* = \sin^{-1}\left(\tfrac{\nu}{\omega r_A}\right), \tag{34}$$
*where $\hat{u}_A^*$ is measured CW w.r.t. the line from $A$ to $T$.*

*Proof.* This result comes from the solution given in [16, Theorem 1] (i.e., the solution to the Turret-Attacker game of min/max terminal distance, which is identical to this game when $D$ is not involved). $A$'s strategy has been slightly modified to account for the assumption that $u_T = 1$. ∎

Finally, the equilibrium capture point for capture by $T$ alone is

$$\mathbf{p}_T^* = \left(r_{\mathbf{p}_T}^*, \theta_{\mathbf{p}_T}^*\right) = \left(r_{A_f}^*, \frac{\omega(r_A - r_A^*)}{\nu}\right), \quad (35)$$

where $r_{A_f}^*$ is defined in (33). Note that $g$ is transcendental and so $g^{-1}$ does not have an analytic form thus $r_{A_f}^*$ must be obtained numerically.

### C. Simultaneous Capture

In the third and final termination case, $A$ is captured by $T$ and $D$ simultaneously; the engagement end with $T$ aligning its line-of-sight with $A$ and $D$ who are coincident. This generally occurs when $A$ is 'between' $T$ and $D$ while not being too close to one or the other.

Because of Lemma 1, the equilibrium capture point for simultaneous capture must occur on the boundaries of both of the previously defined dominance regions; moreover, if multiple intersections exist, it should be the point closest to the origin, i.e.,

$$A_f^* = \min_r \left\{(r,\theta) \in \partial \mathcal{R}_{A/T} \cap \partial \mathcal{R}_{A/D}\right\}. \quad (36)$$

Concerning the computation of $\partial \mathcal{R}_{A/T} \cap \partial \mathcal{R}_{A/D}$, it is straightforward to express the Apollonius circle (i.e., $\partial \mathcal{R}_{A/D}$) in polar coordinates via

$$r = r_\mathbf{c} \cos(\theta - \theta_\mathbf{c}) \pm \sqrt{\rho^2 - r_\mathbf{c}^2 \sin^2(\theta - \theta_\mathbf{c})}, \\ \theta \in \left[\theta_\mathbf{c} - \sin^{-1}\frac{\rho}{r_\mathbf{c}}, \theta_\mathbf{c} + \sin^{-1}\frac{\rho}{r_\mathbf{c}}\right], \quad (37)$$

where $r_\mathbf{c} = \|\mathbf{c}\|$ and $\theta_\mathbf{c} = \operatorname{atan2}(x_\mathbf{c}, y_\mathbf{c})$. Then the intersection closest to the origin may be found by equating the negative versions of (37) and (24) and solving for $\theta$. Define

$$t(\theta) = r_A \cos(\theta - \theta_A) - \sqrt{\frac{\nu^2 \theta^2}{\omega^2} - r_A^2 \sin^2(\theta - \theta_A)}, \\ d(\theta) = r_\mathbf{c} \cos(\theta - \theta_\mathbf{c}) - \sqrt{\rho^2 - r_\mathbf{c}^2 \sin^2(\theta - \theta_\mathbf{c})}, \quad (38)$$

which represent the 'bottom halves' of $\partial \mathcal{R}_{A/T}$ and $\partial \mathcal{R}_{A/D}$, respectively. However, solving for $\theta$ s.t. $t(\theta) = d(\theta)$ cannot be accomplished in closed form and thus the equilibrium capture point must be obtained numerically. Since both $t$ and $d$ have specified domains, the equilibrium capture point must occur in the intersection of these domains, i.e.,

$$\theta_{A_f}^* \in \left[\max\left\{\underline{\theta}, \theta_\mathbf{c} - \sin^{-1}\frac{\rho}{r_\mathbf{c}}\right\}, \min\left\{\overline{\theta}, \theta_\mathbf{c} + \sin^{-1}\frac{\rho}{r_\mathbf{c}}\right\}\right], \\ \text{s.t. } d\left(\theta_{A_f}^*\right) = t\left(\theta_{A_f}^*\right). \quad (39)$$

Finally, the Value of the game, in this case is

$$V(\mathbf{x}) = d\left(\theta_{A_f}^*\right) - 1, \quad (40)$$

and the equilibrium capture point is

$$\mathbf{p}_{DT}^* = \left(r_{\mathbf{p}_{DT}}^*, \theta_{\mathbf{p}_{DT}}^*\right) = \left(d\left(\theta_{A_f}^*\right), \theta_{A_f}^*\right), \quad (41)$$

where $\theta_{A_f}^*$ is defined in (39).

## V. FULL SOLUTION

The previous section established the equilibrium control for $D$ and $A$ and equilibrium outcomes for each of the three termination cases. This section completes the solution through the determination of which termination case is optimal and the equilibrium turning direction for $T$.

**Lemma 7.** *For $\theta_A < \theta_B(r_A)$, the point $\mathbf{p}_D^* \in \mathcal{R}_{A/T}$ if*

$$\|\mathbf{p}_D^*\| > \frac{\nu}{\omega}, \quad (42)$$

*and*

$$\frac{\|A - \mathbf{p}_D^*\|}{\nu} \leq \frac{\operatorname{atan2}(y_{\mathbf{p}_D}^*, x_{\mathbf{p}_D}^*)}{\omega}. \quad (43)$$

*Proof.* The result follows from comparing the time-to-go of $A$ and $T$ to the point $\mathbf{p}_D^*$ which is valid because the assumptions preclude $A$ from passing through the region where it has an angular rate advantage over $T$ (i.e., $r < \frac{\nu}{\omega}$). ∎

**Lemma 8.** *For $\theta_A < \theta_B(r_A)$, the point $\mathbf{p}_T^* \in \mathcal{R}_{A/D}$ if*

$$\|\mathbf{p}^\dagger\| > \frac{v}{\omega}, \quad (44)$$

*and*

$$\frac{\|A - \mathbf{p}^\dagger\|}{\nu} > \frac{\operatorname{atan}(y_\mathbf{p}^\dagger, x_\mathbf{p}^\dagger)}{\omega}, \quad (45)$$

*where $\mathbf{p}^\dagger$ is the point on $\partial \mathcal{R}_{A/D}$ corresponding to $A$ taking the equilibrium $A$ vs. $T$ heading, (34).*

*Proof.* The proof is similar to that of the previous Lemma and is omitted. ∎

Note that the point $\mathbf{p}^\dagger$ can be obtained from geometry and is thus easier to compute than $\mathbf{p}_T^*$ which requires numerical computation of $r_{A_f}^*$.

**Theorem 1** (Termination Determination). *For the differential game specified by (1) – (6) with $T$ turning CCW and $\theta_A < \theta_B(r_A)$ the following hold:*

1) *if the closest point on the Apollonius circle to $T$ is inside $A$'s dominance region w.r.t. $T$, i.e., $\mathbf{p}_D^* \in \mathcal{R}_{A/T}$ then solo capture by $D$ is optimal (where $\mathbf{p}_D^*$ is defined in (18)) and the solution is given by Lemma 2*
2) *if the closest point on $\partial \mathcal{R}_{A/T}$ to $T$ is inside the Apollonius circle, i.e., $\mathbf{p}_T^* \in \mathcal{R}_{A/D}$ then solo capture by $T$ is optimal (where $\mathbf{p}_T^*$ is defined in (35)) and the solution is given by Lemma 6*
3) *otherwise, simultaneous capture by $T$ and $D$ is optimal and all agents head to $\mathbf{p}_{DT}^*$ (defined in (41)) and the Value of the game is $V(\mathbf{x}) = \|\mathbf{p}_{DT}^*\| - 1$.*

*Proof.* For the first two statements, i.e., those concerning solo capture, the points $\mathbf{p}^*$ are the closest point on $A$'s dominance region w.r.t. the capturing agent. If that point is inside of $A$'s dominance region w.r.t. the other agent, it implies, by construction, that capture at that point must occur before the other agent can catch $A$. $A$ can do no better w.r.t. the capturing agent – thus any deviation from heading to $\mathbf{p}^*$ would allow the capturing agent to capture $A$ even further away from $T$.

If neither of the solo capture aim points are safely reachable by $A$, then an intersection between $\partial \mathcal{R}_{A/T}$ and $\partial \mathcal{R}_{A/D}$ must exist by construction. Moreover, this intersection must be the equilibrium capture point (i.e., the closest point in $\mathcal{R}_{A/T} \cap \mathcal{R}_{A/D}$ to $T$) due to the fact that $d(\theta)$ (as defined in (38)) must be monotonically increasing for $\theta > \theta^*_{\mathbf{p}_{DT}}$ and $t(\theta)$ must be monotonically increasing for decreasing $\theta < \theta^*_{\mathbf{p}_{DT}}$. ■

For cases where $\theta_A > \theta_B(r_A)$ the geometry of $\mathcal{R}_{A/T}$ is not well defined. Here, the game (with $T$ turning CCW) may be solved by computing

$$\mathbf{p}^* = (x^*_\mathbf{p}, y^*_\mathbf{p}) = \underset{\mathbf{p} \in \mathcal{F}}{\arg\min} \|\mathbf{p}\| \quad (46)$$

where the feasible set is defined by

$$\mathcal{F} = \left\{ (x,y) \in \mathcal{R}_{A/D} - \mathcal{S} \,\bigg|\, \frac{\|A - (x,y)\|}{\nu} \leq \frac{\operatorname{atan2}(y,x)}{\omega} \right\}, \quad (47)$$

where $\mathcal{S}$ is the target set ($r \leq 1$) and its shadow w.r.t. $A$. The exclusion of $\mathcal{S}$ precludes consideration of possible points in $\mathcal{R}_{A/D}$ wherein $A$'s path would have to pass through the target.

**Theorem 2** (Full Solution). *For the differential game specified by (1) – (6) denote the solution of the game with $T$ turning CCW as $V^+(\mathbf{x})$. When $\theta_A < \theta_B(r_A)$, $V^+$ is given by Theorem 1, otherwise $V^+ = \|\mathbf{p}^*\| - 1$ where $\mathbf{p}^*$ is defined in (46). Denote the solution of the game with $T$ turning CW as $V^-(\mathbf{x})$ which can be obtained by using the same process as for $V^*$ but with $A$ and $D$ mirrored about $T$'s look angle. The solution of the game is*

$$V(\mathbf{x}) = \max\{V^+, V^-\}, \quad (48)$$

*along with the respective equilibrium capture point and agents' strategies associated with the maximizer.*

*Proof.* From Lemma 1 we have that $T$'s control, $u_T$, is constant and is either $-1$ or $1$. The quantities $V^+$ and $V^-$ are the equilibrium Values for $T$ turning CCW or CW in the sense that $A$ and $T$'s strategies are in equilibrium. One of these quantities must be the Value of the game since there are only two options for $T$'s equilibrium control. Therefore, the Value must be the maximum of these since $T$ wishes to capture $A$ as far away as possible. ■

It is possible for the CCW and CW Values to be the same, i.e., $V^+ = V^-$. In this case, the state of the system lies on the Dispersal Surface (DS). Both choices for $T$'s turning direction are equally optimal and thus $T$ is free to choose either one.

It is assumed that this choice may be communicated with $D$ as they are on the same team, and thus $D$ would know the proper equilibrium capture point to aim at. $A$ on the other hand can only guess at which direction $T$ will start turning. If $A$ guesses incorrectly it will suffer a small performance penalty (i.e., be captured further away) and must readjust accordingly. It is shown in the Appendix that the $\mathcal{R}_{A/T}$ corresponding to the direction $T$ turns remains a subset of $\mathcal{R}_{A/T}$ at initial time, and likewise the $\mathcal{R}_{A/T}$ at initial time for the opposite direction remains a subset of all subsequent $\mathcal{R}_{A/T}$ for that direction. Thus $A$ has no way to exploit the singularity (c.f., e.g., [17]).

Examples for each of the three termination cases are shown in Fig. 2 – Fig. 4. In each of the examples $\partial \mathcal{R}_{A/D}$ is blue, $\partial \mathcal{R}_{A/T}$ is green, $\mu = \omega = 1$, and $\nu = 0.7$. Only the CCW turning direction for $T$ is considered as $\theta_A$ is relatively small in each case. The Value of the game is depicted by a dashed purple ring. Initial conditions are shown with filled circles whereas open circles denote terminal conditions.

## VI. CONCLUSION

A novel differential game was formulated and solved involving a cooperative team comprised of a Turret and Defender against an Attacker. The analysis focused on the case wherein the $D$-$T$ team could capture $A$ before it could reach $T$. Three terminal cases are possible: solo capture by $D$, solo capture by $T$, and simultaneous capture by both $D$ and $T$. Conditions were given by which the equilibrium termination could be ascertained from the initial condition and problem parameters. The case wherein $A$ is inside the barrier (i.e., able to reach the speed ratio circle w.r.t. to $T$) was given a cursory treatment in this initial study. A more detailed analysis of the outcome in this case is left for future work. Also, the partitioning of the state space into regions of $A$ winning, solo capture, and simultaneous capture, is left for future work. Most importantly, this work successfully demonstrated an example of cooperation among a heterogeneous team.

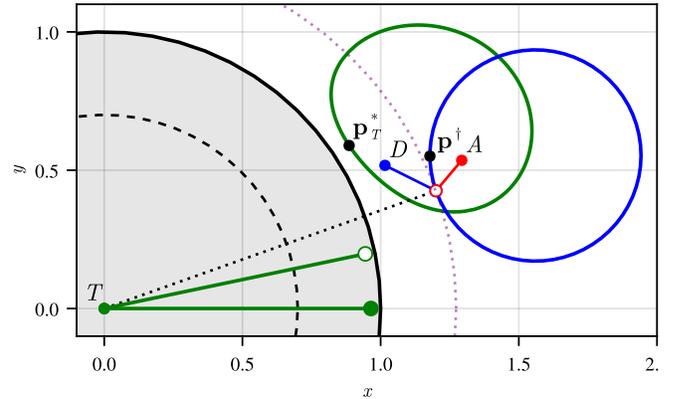

Fig. 2. Solo capture by the Defender at $\mathbf{p}^*_D$.

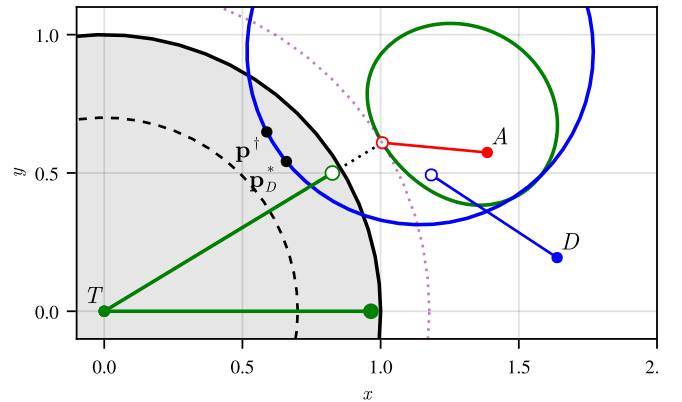

Fig. 3. Solo capture by the Turret at $\mathbf{p}^*_T$.

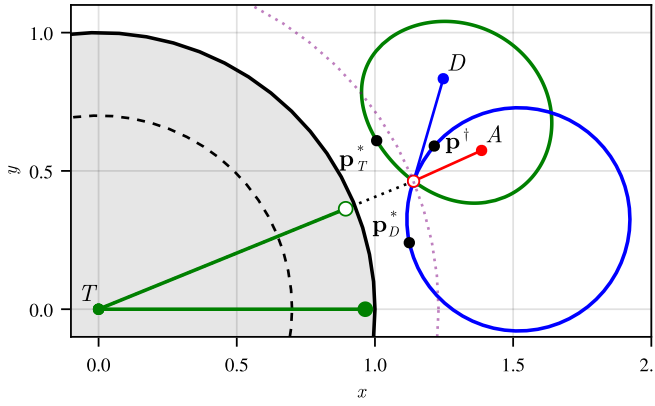

Fig. 4. Simultaneous capture.

## APPENDIX

### A. Monotonicity of Dominance Regions w.r.t. Turret

Suppose, w.l.o.g., that $T$ turns CCW, i.e., $u_T = 1$. Let $\mathcal{R}^+_{A/T}(t)$ denote $A$'s dominance region over $T$ at time $t$ if $T$ were to continue moving CCW. Similarly define $\mathcal{R}^-_{A/T}(t)$ as if $T$ were to switch to CW, i.e., $u_T = -1$. In the following, it is shown that $\mathcal{R}^+_{A/T}(t)$ strictly contracts while $\mathcal{R}^-_{A/T}(t)$ strictly expands.

**Lemma 9**. *If $T$ turns CCW and $\theta_A < \theta_B(r_A)$ then*
$$\mathcal{R}^+_{A/T}(t_2) \subset \mathcal{R}^+_{A/T}(t_1), \tag{49}$$
*for all $t_2 > t_1 \geq 0$.*

*Proof.* (by contradiction). Let there exist a point $\mathbf{p} \in \mathcal{R}^+_{A/T}(t_2)$ such that $\mathbf{p} \notin \mathcal{R}^+_{A/T}(t_1)$. Since $\mathbf{p} \in \mathcal{R}^+_{A/T}(t_2)$, $A$ can reach $\mathbf{p}$ before getting captured by $T$. That is,
$$\exists t_f > t_2 \text{ such that } A(t_f) = \mathbf{p}. \tag{50}$$
Let $A(t)$ for $t \in [t_2, t_f]$ denote the corresponding trajectory of $A$. Now consider $A$'s concatenated trajectory, $A(t)$ for $t \in [t_1, t_f]$, which shows that $A(t_f) = \mathbf{p}$ when $T$ is moving CCW for the entire duration $[t_1, t_f]$. Consequently, $\mathbf{p}$ must be in $\mathcal{R}^+_{A/T}(t_1)$. ■

**Lemma 10**. *If $T$ turns CCW and $2\pi - \theta_A < \theta_B(r_A)$ then*

$$\mathcal{R}^-_{A/T}(t_1) \subset \mathcal{R}^-_{A/T}(t_2), \tag{51}$$

*for all $t_2 > t_1 \geq 0$ where $A$ has not been captured or reached its target prior to $t_2$.*

*Proof.* The proof is based on comparing the times to go for $A$ and $T$ to a point, which is how $\mathcal{R}_{A/T}$ is constructed. Recall that $T$ moves CCW whereas $\mathcal{R}^-_{A/T}(t)$ is $A$'s dominance region w.r.t. $T$ if $T$ were to move *clockwise* starting at $t$. Consider a point $\mathbf{p} \in \mathcal{R}^-_{A/T}(t_1)$ – the point $\mathbf{p}$ is reachable by $A$ without being captured when $T$ moves CW. Consider any possible trajectory of $A$ in the time interval $t \in [t_1, t_2]$. That is, $A(t_2)$ may be anywhere in the disk of radius $\nu(t_2 - t_1)$ centered on $A(t_1)$. In the meantime, it has been assumed that $T$ moved CCW.

Now it is shown that $\mathbf{p}$ must be in $\mathcal{R}^-_{A/T}(t_2)$. If $T$ were to, at $t_2$ begin moving CW, by the time that $T$ reached its look angle at $t_1$, (i.e., $T(t_1)$) $A$ can reach $A(t_1)$ since $\|A(t_2) - A(t_1)\| \leq \nu(t_2 - t_1) = \frac{\nu}{\omega}(T(t_2) - T(t_1))$ by construction – hence $\mathbf{p} \in \mathcal{R}^-_{A/T}(t_2)$ since the agents have returned to their initial positions. ■

### B. Computation of $\mathbf{p}^\dagger$

The Apollonius circle represents the locus of points where $A$ is captured by $D$ when $A$ takes a constant heading and $D$ is on the collision course. In other words, it is the locus of points whose ratio of distances to $A$ and $D$ are equal to the speed ratio $\frac{\nu}{\mu}$. First, from [18], the distance traveled by $A$ to a point $\mathbf{p}$ on the Apollonius circle along a constant heading $\psi$ measured w.r.t. the line of sight $A - D$ is given by

$$\|A - \mathbf{p}\| = \alpha \|A - D\| \left( \cos \psi + \sqrt{\frac{\mu^2}{\nu^2} - \sin^2 \psi} \right). \tag{52}$$

The point $\mathbf{p}^\dagger$ is based on $A$ taking its 1v1 equilibrium heading w.r.t. $T$ which is [16]

$$\hat{u}_A = \sin^{-1}\left(\frac{\nu}{\omega r_A}\right). \tag{53}$$

Converting to the Cartesian frame gives
$$u_A = \theta_A - \pi - \hat{u}_A. \tag{54}$$
Finally, converting this heading to an angle relative to the line-of-sight gives
$$\psi^\dagger = u_A - \operatorname{atan2}(y_A - y_D, x_A - x_D). \tag{55}$$
Substituting this quantity in for $\psi$ into (52) gives the distance $\|A - \mathbf{p}^\dagger\|$. Finally,
$$\mathbf{p}^\dagger = \|A - \mathbf{p}^\dagger\| \begin{bmatrix} \cos \psi^\dagger \\ \sin \psi^\dagger \end{bmatrix}. \tag{56}$$